\documentstyle[a4]{article}

\def\laq{\mathrel{\mathchoice{\vcenter{\offinterlineskip\halign{\hfil
$\displaystyle##$\hfil\cr<\cr=\cr}}}
{\vcenter{\offinterlineskip\openup
0,5mm\halign{\hfil$\textstyle##$\hfil\cr<\cr=\cr}}}
{\vcenter{\offinterlineskip\halign{\hfil$\scriptstyle##$\hfil\cr<\cr=\cr}}}
{\vcenter{\offinterlineskip\halign{\hfil$\scriptscriptstyle##$\hfil\cr<\cr=
\cr}}}}}
\def\gaq{\mathrel{\mathchoice{\vcenter{\offinterlineskip\halign{\hfil
$\displaystyle##$\hfil\cr>\cr=\cr}}}
{\vcenter{\offinterlineskip\openup
0,5mm\halign{\hfil$\textstyle##$\hfil\cr>\cr=\cr}}}
{\vcenter{\offinterlineskip\halign{\hfil$\scriptstyle##$\hfil\cr>\cr=\cr}}}
{\vcenter{\offinterlineskip\halign{\hfil$\scriptscriptstyle##$\hfil\cr>\cr=
\cr}}}}}

\renewcommand{\theequation}{\mbox{\arabic{section}.\arabic{equation}}}
\renewcommand{\thefootnote}{\fnsymbol{footnote}}

\begin{document}

\begin{titlepage}
{}
\vskip 2cm
\centerline{\bf CHIRAL AND CONTINUOUS SYMMETRY}
\vskip 0,5cm
\centerline{\bf OF AN $XY$ SPIN GLASS}
\vskip 0,5cm
\centerline{\bf ON A TUBE LATTICE}
\vskip 4cm
\renewcommand{\thefootnote}{\fnsymbol{footnote}}
\centerline{M.J.~Thill, M.~Ney-Nifle\footnote[2]{\noindent Permanent address:
Laboratoire de Physique\footnotemark[1], Ecole Normale Sup\'erieure, F-69364
Lyon CEDEx 07} and H.J.~Hilhorst}
\vskip 0,3cm
\centerline{Laboratoire de Physique Th\'eorique et Hautes Energies
\footnote[1]{\noindent Laboratoire associ\'e au Centre National de la Recherche
Scientifique}}
\vskip 0,2cm
\centerline{B\^atiment 211, Universit\'e de Paris-Sud, F-91405 Orsay CEDEx}
\vskip 5cm

{\centerline \today}

\vskip 5cm

{\noindent\bf LPTHE Orsay 94/23}
\vskip 0,3cm

\end{titlepage}

\begin{abstract}
\thispagestyle{empty}

We analyse the chiral symmetry in the random $\pm J$ $XY$ model on a
$N\times 2$ square lattice with periodic boundary conditions in the
transverse direction. This ``tube" lattice may be seen as a
two-dimensional lattice of which one dimension has been compactified.
In the Villain formulation the discrete-valued {\em chiralities}\/
or {\em charges}\/ associated with the plaquettes of the lattice
decouple from the continuous degrees of freedom. The difficulty
of the problem lies in the fact that the chiralities interact
through the long range ``strong" one-dimensional Coulomb potential - which
increases linearly with distance - as well as through an
exponentially decaying ``weak" interaction. By comparing the
ground state energies for periodic, antiperiodic, and reflecting
boundary conditions in the longitudinal direction, we show that
the chiralities and the $XY$ spins have the {\em same}\/ zero-$T$
correlation length exponent, whose exact value $\nu_c = 0.5564\ldots$
we determine. The equality of these correlation lengths even in the
presence of long range chirality-chirality interactions lends support
to the view that chiral glass order cannot be sustained without
simultaneous spin glass order.

\end{abstract}
\vfill\eject

\setcounter{page}{1}

\section{Introduction}

We present a study of the interplay between the spin variables and
the chiral variables (chiralities) in the $\pm J$ $XY$ spin glass.
The former correspond to the continuous rotational symmetry of this
model, and the latter to its discrete
chiral symmetry (i.e., the invariance of the model Hamiltonian under reflection
of all the spins with respect to a reference axis), first pointed out by
Villain \cite{Vf,Vspg,Vd3}. Below the lower critical dimension, $d_\ell$, which
is believed to be greater than $2$ {lcd}, the
correlation lengths associated with the chiralities and with the spin
variables diverge as $T^{-\nu_c}$ and $T^{-\nu_s}$, respectively, at
the zero-temperature ($T$) critical point. The question of the
relation between the two types of variables has become of interest following
speculations by Kawamura and Tanemura \cite{KaT}, by Ray and Moore
\cite{RM} and by Kawamura \cite{Ka}, prompted
by Monte Carlo simulations, that below $d_\ell$, the two correlation lengths
are different, with $\nu_c>\nu_s$. This suggests that the chiralities
will order more easily than the spins in higher dimensions. Consequently, above
$d_\ell$ there would be a regime of
dimensions with long range chiral glass order, but without conventional spin
glass
order. This possibility receives intuitive support from the idea that
discrete symmetry leads to long range order more easily than continuous
symmetry does.

Two recent publications \cite{NHM,NH} address this issue analytically. Both
these
studies, just like the Monte Carlo work of \cite{KaT,RM}, consider the finite
size scaling of the ground state energy differences between periodic (P),
antiperiodic (AP), and reflecting (R) boundary conditions. In one of them,
Ney-Nifle and
Hilhorst \cite{NH} transform the two-dimensional $XY$ $\pm J$ spin glass on a
finite
$N\times M$ square lattice into a grand-canonical Coulomb gas problem of which,
as is well-known, the logarithmically interacting charges represent the chiral
variables.
The charges must take half-integer values on the frustra\-ted plaquettes and
therefore cannot vanish even in the ground state. In the case of uncorrelated
disorder, the plaquettes are randomly and independently frustrated with
probabi\-lity $\frac{1}{2}$, and it is not possible to find the ground state
explicitly.
For that reason, the subsequent analytic treatment of \cite{NH} remains
restricted to the example of a rectan\-gular array of frustrated plaquettes
with
randomly distributed intercolumn distances. In this example, the authors find
no evidence for
a chiral correlation length diverging faster than the spin correlation length.
By a heuristic argument they extend this conclusion to the case of uncorrelated
$\pm J$ disorder.

In an earlier investigation, Ney-Nifle {\em et al}\/ {NHM} considered the
random $\pm J$
$XY$ model on a one-dimensional ladder lattice, again in the Coulomb gas
representation. This problem is exactly solvable, or nearly so, for general
disorder, and the conclusi\-ons drawn from it are fully coherent with those
from the two-dimensional model \cite{NH}. However, this model suffers from the
drawback
that, in the Coulomb gas language, it has only exponentially decaying
electrostatic interactions (for reasons explained in that work), so that one
may
wonder if an essential ingredient of the difficult two-dimensional problem
has not been lost.

In the present work, we reconcile the requirements of exact solvability and
truly
long-range interactions between the chiralities by studying the $\pm J$ $XY$
spin glass on a $N\times 2$
lattice which is periodic both in the longitudinal and the transverse
direction. We work again in the Coulomb gas representation, and apply different
boundary conditions. In section 2, we show that, on this two-dimensional
lattice with one compactified dimension, the electrostatic
interaction decomposes into two components. The first one is
a ``strong" or charge-charge interaction; it is nothing but the one-dimensional
Coulomb potential, which increases linearly with distance.
The second one is a ``weak" interaction: it acts between transversely
oriented ``dipoles" and decays exponentially with distance. We shall refer
to them as the Coulomb and the dipolar interaction, respectively. The
Coulomb gas representation of the $XY$ model Hamiltonian involves, in addition
to these
two interactions, two supplementary ``global" terms that
couple the system's total electric dipole moment to the boundary conditions
imposed on the Hamiltonian. These extra terms have drawn a certain attention in
the recent literature \cite{FTY,NHM,NH}, and they play again an important
r\^ole here.

We are not able to solve the ground state problem for the complete
Hamiltonian. However, we are able to conclude that in the large $N$ limit,
whatever the boundary conditions,
the ground state is one of the infinitely many ground states of the
``strong" Coulomb interaction combined with one of the global terms. The
details of the proof (of largely technical nature) of this fact are given in
Appendix A. This set of ground states consists, roughly speaking, of charge
configurations
in which the long-range Coulomb interaction is screened away as much as
possible by the formation of longitudinally oriented dipoles, as exhibited in
section 3. The
degeneracy within this set is lifted by the weak interaction and by the second
global term, which are therefore responsible for the selection of the ground
state of the full Hamiltonian and for the energy
differences between P, AP, and R boundary conditions. Even though we
remain unable to say which member of this set is selected as the true
ground state, we are able to describe (in section 4) the
domain walls and domain wall energies involved in passing from one
boundary condition to another. Using the relation between the correlation
lengths and the finite size scaling exponent of the ground state energy
differences, we conclude in the final section 5, for the first time within
an $XY$ spin glass with random $\pm J$ disorder and having a nontrivial
long-range interaction between its chiralities, that the spin and the
chiral correlation lengths diverge, for $T \to 0$, with the {\em same}\/
exponent $\nu$. We determine its exact value,
$\nu=\log^{-1}_{\frac{8}{3}}(3+2\sqrt{2})=0.5564\ldots$, in Appendix B.

\section{The tube: a compactified two-dimensional lattice}

In this section we shall exhibit the Hamiltonians of a random $XY$ model
on a $N\times 2$ square lattice with periodic boundary conditions (PBC)
in the transverse direction and successively periodic, antiperiodic, and
reflecting boundary conditions in the other direction. We call this
lattice a {\em tube}\/ (see figure 1). It can be viewed as a
two-dimensional lattice of which one dimension has been compactified.

First we shall recall the same model on the more general $N\times
M$ lattice and then specialise to the tube. The effect of the
compactification on the interaction will thereby become clear.

We consider a random $\pm J$ $XY$ model where the spins are two-component
unit vectors whose angles ${\phi_i}$ (with a reference axis) take values
in $(-\pi, \pi]$. Two nearest-neighbour spins, $\phi_i$ and $\phi_j$,
have an interaction energy $-J\cos(\phi_i-\phi_j-\pi_{ij})$, where $J$ is a
constant, and the $\pi_{ij}$ are quenched random variables that take the values
\begin{equation}
\pi_{ij} = \left\{ \begin{array}{r@{\quad \quad}l} 0 \, , & \mbox{\small
with probability} \, \, {1 \over 2} \\ \pi \, , & \mbox{\small with
probability} \, \, {1 \over 2}  \end{array} \right. .
\end{equation}
The partition function is
\begin{equation}\label{ZXY}
Z_{XY} = \int\limits^\pi_{-\pi} \prod_{i} d\phi_{i}
\,\, \mbox{e}^{\beta J \sum\limits_{<i,j>} \cos(\phi_i - \phi_j -
\pi_{ij})}\,\, .
\end{equation}
The sum in the exponential in (\ref{ZXY}) runs over all
nearest-neighbour bonds of the periodic lattice with the convention
that in $<i,j>$ the site $j$ is to the right of $i$ (for a horizontal
bond) or above $i$ (for a vertical bond). In our notation the site vectors
$i=(i_x,i_y)$ have half-integer components
$i_x=\frac{1}{2},\frac{3}{2},\ldots,\frac{2N-1}{2}$ and
$i_y=\frac{1}{2},\frac{3}{2},\ldots,\frac{2M-1}{2}$.
\begin{figure}
\setlength{\unitlength}{1cm}
\begin{picture}(13,6)
\thicklines
\put(1.5,1){\line(1,0){11}}
\put(1.5,3){\line(1,0){11}}
\put(1.5,5){\line(1,0){0.125}}
\put(12.375,5){\line(1,0){0.125}}
\multiput(1.875,5)(0.5,0){21}{\line(1,0){0.25}}
\multiput(2,1)(2,0){6}{\line(0,1){4}}
\thinlines
\put(1.9,2){\line(1,0){0.2}}
\put(1.9,4){\line(1,0){0.2}}
\put(3,0.9){\line(0,1){0.2}}
\put(5,0.9){\line(0,1){0.2}}
\put(7,0.9){\line(0,1){0.2}}
\put(9,0.9){\line(0,1){0.2}}
\put(11,0.9){\line(0,1){0.2}}
\put(0.75,1.25){\vector(0,1){1.5}}
\put(0.75,1){\makebox(0,0){$y$}}
\put(12,0.4){\vector(1,0){1.5}}
\put(11.75,0.4){\makebox(0,0){$x$}}
\put(1.5,2){\makebox(0,0){$1$}}
\put(1.5,4){\makebox(0,0){$2$}}
\put(9,0.7){\makebox(0,0){$N$}}
\put(11,0.7){\makebox(0,0){$1$}}
\put(3.5,2){\makebox(0,0){$\pi_{x_0}^{12}$}}
\put(6.5,2){\makebox(0,0){$\pi_{x_0+1}^{12}$}}
\put(5,3.25){\makebox(0,0){$\pi_{(x_0,2)}$}}
\put(5,1.25){\makebox(0,0){$\pi_{(x_0,1)}$}}
\put(3.5,4){\makebox(0,0){$\pi_{x_0}^{21}$}}
\put(6.5,4){\makebox(0,0){$\pi_{x_0+1}^{21}$}}
\put(5,0.6){\makebox(0,0){$x_0$}}
\thicklines
\put(5,2){\makebox(0,0){\boldmath $p_{(x_0,1)}$}}
\put(5,4){\makebox(0,0){\boldmath $p_{(x_0,2)}$}}
\end{picture}

{\footnotesize {\noindent\bf\footnotesize Figure 1:} {\sl\footnotesize
Chiralities on a tube lattice.}  The bonds on the dashed line
are identical with the ones on the lower solid line. The bonds have quenched
disorder variables $\pi$ and the plaquette centres frustration variables $p$
associated with them. According to the definition (2.4),
one has, for example, $p_{(x_0,1)}=(\pi_{(x_0,1)} + \pi_{x_0+1}^{12} -
\pi_{(x_0,2)} - \pi_{x_0}^{12})/(2\pi)$.}
\end{figure}

Since we are interested in the ground state properties of the model, we
shall replace (\ref{ZXY}) by the corresponding Villain expression, which is
believed to be equivalent to
(\ref{ZXY}) in the large-$\beta$ limit \cite{Vf,Vspg}, and is easier to
analyse. The Villain
partition function is
\begin{equation}\label{ZV}
Z_{\mbox{\tiny V}} = \int^\pi_{-\pi} \prod_i d\phi_i
\sum_{\{n_{ij}\}} \mbox{e}^{-\frac{\beta J}{2} \sum\limits_{<i,j>}(\phi_i -
\phi_j - \pi_{ij}
- 2\pi n_{ij})^2}\,\, ,
\end{equation}
where the $n_{ij}$ are additional dynamical variables. These $n_{ij}$
are integers and the sum on them ensures that the integrand has period
$2 \pi$ in $\phi_i - \phi_j$. In the following, we set $J=2$.

For each plaquette of the lattice, we define a frustration
variable $p_r$, with $r = (x,y)$ a vector with integer components
$x=1,\ldots,N$ and $y=1,\ldots,M$ that labels the centres of the plaquettes,
\begin{equation}\label{pr}
p_r \equiv {\sum\limits_{<i,j>}\!}^{(r)}\, \epsilon_{ij}^r\,\,
\frac{\pi_{ij}}{2\pi}\,\, ,
\end{equation}
where the sum is restricted to the bonds that define the plaquette $r$.
In (\ref{pr}), $\epsilon_{ij}^r = -1$ or $1$ depending on whether
one runs through the
corners of the triangle $(ijr)$ clockwise or counterclockwise.
The frustration variable
is {\em integer}\/ for {\em nonfrustrated}\/
plaquettes and {\em half-integer}\/ otherwise.

In (\ref{ZV}) one can integrate on the continuous degrees of freedom.
The algebra (cf.~\cite{Vf,Vspg,NHM,NH}) includes
the transformation
from the variables
$n_{ij}$ to the new discrete variables $q_r$ called the ``chiralities''
of the plaquettes. The chirality $q_r$ runs through all integers
(half-integers) when $p_r$ is integer (half-integer). One shows that
the chiralities interact via a Coulomb interaction
(which is why they are also called ``charges'') and that they satisfy
the neutrality condition
\begin{equation}
\sum\limits_r q_r = 0\,\, .
\end{equation}

Recently, Ney-Nifle and Hilhorst \cite{NH} (see also \cite{NHM}) extended the
mapping of the
$XY$ Hamiltonian onto a Coulomb gas Hamiltonian by including all the
finite size corrections on a $N\times M$ lattice with various
boundary conditions. We will now adapt their results to the tube
lattice, for which a simplified notation is defined in figure 1.

\subsection{Periodic boundary conditions}

We shall first consider the $N\times 2$ system with periodic boundary
conditions [PBC] in the longitudinal direction. We denote its
partition function by $Z_{\mbox{\tiny P}}$. Starting from the more general
model $Z_{\mbox{\tiny V}}$ \cite{NH},
see equation (\ref{ZV}), we change variables from $n_{ij}$ to the
chiralities $q_r$ which allows to perform the Gaussian integration
on the first set of variables, $\phi_i$. Including all numerical
prefactors in $Z_0^{\mbox{\tiny P}}$, one gets \cite{NH}
\begin{equation}\label{ZP}
Z_{\mbox{\tiny P}} = Z_0^{\mbox{\tiny P}}\,  \sum_{\{ q_r \} }\, \sum_{n,m}\,
e^{-\beta {\cal H}_{\mbox{\tiny P}}}\,
\delta \left(\sum_r q_r, 0 \right) \,\, ,
\end{equation}
where $\delta(\cdot,\cdot)$ denotes the Kronecker delta.
The additional dynamical variables $n$ and $m$ run over all integers and
the $q_r$ take integer or half-integer values, as mentioned above. The
Hamiltonian ${\cal H}_{\mbox{\tiny P}}$, which will be the starting point of
our considerations, reads explicitly \cite{NH} \begin{equation}\label{HPBC}
\begin{array}{lcl}
{\cal H}_{\mbox{\tiny P}} &=& \hspace{0,5cm}\frac{8\pi^2}{N} \left(n+
\frac{1}{2}
\sum\limits_{x=1}^N q_{(x,1)} + \sum\limits_{x=1}^N
\frac{\pi_{(x,1)}}{2\pi} \right)^2  \\[2mm]
{} &{}& + \,\, 2\pi^2N \left(m+{1 \over N}\sum\limits_{x=1}^N
x(q_{(x,1)}+q_{(x,2)}) +
\frac{\pi_N^{12} + \pi_N^{21}}{2\pi} \right)^2 \\[2mm]
{} &{}& + \,\, \pi^2 \sum\limits_{r,r'} q_rq_{r'}U_{N,2}(r-r') \,\, .
\end{array}
\end{equation}
We will briefly discuss its meaning. The first two terms are due to the
finite system size. They represent a coupling of the horizontal and the
vertical component of the total electric dipole moment, respectively, to the
quenched disorder. In the third term, $U_{N,M}(R)$ is the interaction between
two
charges
\begin{equation}\label{U}
U_{N,M}(R) = \frac{1}{2N}\,{\sum\limits_{k_x,k_y}\!}^*\,
\frac{e^{i(X k_x + Y k_y)}-1} {\sin^2(\frac{k_x}{2}) +
\sin^2(\frac{k_y}{2})}\,\, ,
\end{equation}
with $R=(X,Y)$, $k_x = 0, \frac{2\pi}{N},\ldots ,
\frac{2\pi(N-1)}{N}$ and $k_y =
0, \frac{2\pi}{M},\ldots , \frac{2\pi(M-1)}{M}\, $. The asterisk
indicates that the term $(k_x,k_y)=(0,0)$ is left
out of the summation.

In $d=2$,
$U_{N,M}$ ($N,M\to\infty$) is the two-dimensional Coulomb interaction which
varies as a
logarithm at large distances \cite{Vf,Vspg}. For the tube, we will see in what
follows that the compactification leads to a decomposition of $U_{N,2}$ into
two parts: a one-dimensional Coulomb interaction that increases linearly with
distance and an
exponentially decreasing interaction, which is a remnant of a two-dimensional
dipole-dipole interaction. The appearence of the linear Coulomb interaction
and its
competition with the dipolar interaction makes the model interesting.

To separate these two interactions in $U_{N,2}$, we combine the two
chiralities of a column $x$ as
\begin{equation}\label{q}
\begin{array}{lcc}
q_x^+ &\equiv& q_{(x,1)} + q_{(x,2)} \,\, , \\
q_x^- &\equiv& q_{(x,1)} - q_{(x,2)} \,\, .
\end{array}
\end{equation}
Introducing $q_x^+$ and $q_x^-$ in (\ref{HPBC}) and evaluating (\ref{U})
for $N\to\infty$ in these new variables, we get
\begin{equation}\label{HPBC1}
\begin{array}{lcl}
{\cal H}_{\mbox{\tiny P}} &=& \hspace{0,5cm}\frac{8\pi^2}{N}
\left(n+\frac{1}{4}\sum\limits_{x=1}^N q_x^-
+ \sum\limits_{x=1}^N \frac{\pi_{(x,1)}}{2\pi} \right) ^2 +
\,\, \pi^2 \sum\limits_{x,x'=1}^N q_x^-q_{x'}^-U_{\mbox{\tiny P}}^-(x-x')
\\[2mm]
{} &{}& +\,\, 2\pi^2N \left(m+{1 \over N}\sum\limits_{x=1}^N xq_x^+ +
\frac{\pi_N^{12} + \pi_N^{21}}{2\pi}\right)^2 + \,\, \pi^2
\sum\limits_{x,x'=1}^N q_x^+q_{x'}^+U_{\mbox{\tiny P}}^+(x-x') \,\, .
\end{array}
\end{equation}
We find that the charges $q_x^+$ interact via the long-range periodised Coulomb
potential
\begin{equation}
\begin{array}{lcl}\label{UP+}
U_{\mbox{\tiny P}}^+(X) &=& \hspace{0,5cm} \frac{1}{2N}\,{\sum\limits_{k_x\not=
0}}\,
\frac{e^{i k_x X}-1} {\sin^2(\frac{k_x}{2})}\\[2mm]
{} &\simeq& -\,|X|(1-{{|X|} \over N})\, , \qquad N\to\infty, \,
\frac{|X|}{N}\,\, \mbox{fixed}, \, 0\laq{{|X|} \over
N}\laq 1\,\, .
\end{array}
\end{equation}
If $|X|$ is negligible with respect to $N$, $U^+_{\mbox{\tiny P}}(X)$ is the
usual one-dimen\-sio\-nal Coulomb interaction, linear in $X$. If not, the
term $1-\frac{|X|}{N}$ becomes impor\-tant and reflects the symmetry and
periodicity of the lattice.

The charges $q_x^-$ interact via a short-range (dipolar) potential
\begin{equation}\label{U-P}
\begin{array}{lcl}
U_{\mbox{\tiny P}}^-(x-x') &=& \hspace{0,5cm}
\frac{1}{2N}\,{\sum\limits_{k_x}}\,
\frac{e^{i k_x (x-x')}} {1 + \sin^2(\frac{k_x}{2})}\\[2mm]
{} &\simeq& {\sqrt{2} \over 8} (3-2\sqrt{2})^{d(x,x')}\,\, ,
\qquad N\to\infty ,\,\, |x-x'|\,\, \mbox{fixed},
\end{array}
\end{equation}
where $d(x,x')$ is the length of the shortest path between $x$ and $x'$,
taking into account the periodic geometry.
Furthermore, one obtains from the calculation that both potentials have
the symmetry properties
\begin{equation}
\begin{array}{lcl}
U^{\pm}_{\mbox{\tiny P}}(X) &=& U^{\pm}_{\mbox{\tiny P}}(X+N) \,\, , \\
U^{\pm}_{\mbox{\tiny P}}(X) &=& U^{\pm}_{\mbox{\tiny P}}(-X)  \,\, .
\end{array}
\end{equation}
Because of the range of the interactions, we will also call the
long-range Coulomb interaction between the charges $q_x^+$ {\em strong}\/
interaction and the short-range (dipolar) interaction between the charges
$q_x^-$ {\em weak}\/
interaction. In the large $N$ limit, for convenience, we
rewrite the Coulomb interaction term in (\ref{HPBC1}), using (\ref{UP+}) and
charge neutrality, as
\begin{equation}\label{UP+1}
\pi^2 \sum\limits_{x,x'=1}^N q_x^+q_{x'}^+U_{\mbox{\tiny P}}^+(x-x')\,\, =\,\,
- \pi^2 \sum\limits_{x,x'=1}^N |x-x'|q_x^+q_{x'}^+
- \frac{2\pi^2}{N} \left( \sum\limits_{x=1}^N xq_x^+ \right)^2\,\,  .
\end{equation}
Inserting this expression in (\ref{HPBC1}) and writing out also the
interaction potentials $U_{\mbox{\tiny P}}^-$ explicitly, we obtain eventually
\begin{equation}\label{HPBCf}
\begin{array}{rcl}
{\cal H}_{\mbox{\tiny P}} &=& \hspace{0,5cm}\frac{8\pi^2}{N} \left(n+
\frac{1}{4}
\sum\limits_{x=1}^N q_x^-
+ \sum\limits_{x=1}^N \frac{\pi_{(x,1)}}{2\pi}\right)^2 + \,\,
{\sqrt{2} \over 8}\pi^2 \sum\limits_{x,x'=1}^N
(3-2\sqrt{2})^{d(x,x')}q_x^-q_{x'}^- \\
{} &{}& +\,\, 2\pi^2N \left(m+\frac{\pi_N^{12} + \pi_N^{21}}{2\pi}
\right)^2 -
4\pi^2\left(m+\frac{\pi_N^{12} +
\pi_N^{21}}{2\pi}\right)\sum\limits_{x=1}^N
xq_x^+ - \,\, \pi^2 \sum\limits_{x,x'=1}^N |x-x'|q_x^+q_{x'}^+ \,\, ,
\end{array}
\end{equation}
valid in the large $N$ limit.
The task will now be to minimise ${\cal H}_{\mbox{\tiny P}}$ with respect to
the
four variables $q_x^+$,
$q_x^-$, $m$, and $n$ in order to find its ground state energy. This
will be done in section III.

\subsection {Antiperiodic boundary conditions}

Passing from PBC to antiperiodic boundary conditions (APBC) means
changing the sign of the two horizontal bonds that belong to the plaquettes
$(N,1)$ and $(N,2)$. Under this change frustrated (unfrustrated)
plaquettes remain frustrated (unfrustra\-ted). Thus the only modification
needed to get the Hamiltonian ${\cal H}_{\mbox{\tiny AP}}$ for APBC is to
replace
$\pi_{(N,1)}$ by $\pi_{(N,1)} + \pi$ in the first term in equation
(\ref{HPBCf}), i.e., to add $\frac{1}{2}$ in the expression between
parentheses in that term.

\subsection {Reflecting boundary conditions}

One obtains the Hamiltonian ${\cal H}_{\mbox{\tiny R}}$ for
the $XY$ spin glass on an $N\times M$ lattice with reflecting
boundary conditions (RBC) in the horizontal direction and PBC in the
vertical direction by replacing the horizontal interactions in one single,
but arbitrary column, say $N$, by
\begin{equation}
(\phi_i + \phi_j - 2\pi n_{ij} - \pi_{ij} )^2 \,\, .
\end{equation}
This amounts to reflecting the spins on one side of this column with
respect to the reference axis.
The ensuing modifications in passing to the Coulomb representation result in
\cite{NH}
\begin{equation}\label{ZR}
Z_{\mbox{\tiny R}} = Z_0^{\mbox{\tiny R}}\, \sum_{\{q_r\}}\, \sum_{{n,m}}\,
e^{-\beta {\cal H}_{\mbox{\tiny R}}}\,
\delta\left(\left[\sum_r q_r - {{\pi_N^{12} + \pi_N^{21}} \over \pi}\right]
\bmod 2, 0\right)
\end{equation}
with
\begin{equation}
{\cal H}_{\mbox{\tiny R}} = \pi^2\,\sum_{{r,r'}} q_rq_{r'}U_{\mbox{\tiny
R}}(r-r')\,\, ,
\end{equation}
and $r=(x,y)$ labels the centres of the plaquettes on the $N\times M$
lattice as before. We do not
recall the explicit, general, form for the potential $U_{\mbox{\tiny R}}$ here,
but rather
use the variables $q_x^+$ and $q_x^-$ defined in (\ref{q}) and give
the explicit expression of ${\cal H}_{\mbox{\tiny R}}$ in the case of the tube
lattice:
\begin{equation}\label{HRBC}
{\cal H}_{\mbox{\tiny R}} = \pi^2\sum\limits_{x,x'=1}^N q_x^+q_{x'}^+\,
U_{\mbox{\tiny R}}^+(x-x') +
\pi^2\sum\limits_{x,x'=1}^N q_x^-q_{x'}^-\, U_{\mbox{\tiny R}}^-(x-x')
\end{equation}
where
\begin{equation}\label{UR}
\begin{array}{ccll}
U_{\mbox{\tiny R}}^+(X) &=& {N \over 2} (1-{{2|X|} \over N})\, ,& \quad
N\to\infty, \,
\frac{|X|}{N}\,\, \mbox{fixed, $0\laq{|X| \over
N}\laq 1$}\,\, , \\[2mm]
U_{\mbox{\tiny R}}^-(x-x') &=& {\sqrt{2} \over 8} (3-2\sqrt{2})^{d(x,x')}
\sigma
(x,x')\,\, , &
\quad N\to\infty \,\, , \,\, |x-x'|\,\, \mbox{fixed}\,\, ,
\end{array}
\end{equation}
where $\sigma (x,x') = -1$ if the shortest path between
$x$ and $x'$ crosses the bond $\pi_N^{12}$ (or $\pi_N^{21}$),
and $\sigma (x,x') = 1$ otherwise. The interactions $U_{\mbox{\tiny R}}^+$ and
$U_{\mbox{\tiny R}}^-$
differ furthermore in their symmetry properties from those of PBC in that
one has now
\begin{equation}\label{URS}
U^{\pm}_{\mbox{\tiny R}}(X) = - U^{\pm}_{\mbox{\tiny R}}(X+N) \,\, ,
\end{equation}
i.e., antiperiodicity of the interaction potentials.

\subsection{The Hamiltonians in terms of electric field energy}

In this subsection, we rewrite the strong interaction part of the
Hamiltonians (\ref{HPBCf}) and (\ref{HRBC}) in terms of an electric
field $E_x$:
\begin{equation}\label{E1}
E_x = E_0 + \sum_{x'=1}^x q_{x'}^+\,\, .
\end{equation}
$E_x$ is the electric field between $x$ and $x+1$, and $E_0$ is a constant
background field whose value will be set later in such a way that the volume
sum of the energy density $E_x$ gives the Coulomb energy of the Hamiltonians.
The advantage of this rewriting is clearly seen in Appendix A when properties
of the ground states of the Hamiltonians (for large $N$) are proven: $E_x$ is a
local variable, whereas $\sum\limits_{x'=1}^N q_{x'}^+ U_{\mbox{\tiny
P,R}}^+(x-x')$ involves all columns of the lattice. To determine the effects on
the system's energy when a charge $q_{x_0}^+$ is changed in a given
configuration is much easier in terms of the local variable $E_{x_0}$, as is
manifest in Appendix A.

Squaring (\ref{E1}) and summing over $x$, we get
\begin{equation}
\begin{array}{lcl}\label{E2}
\sum\limits_{x=1}^N E_x^2 &=& NE_0^2+ 2E_0\sum\limits_{x=1}^N
\sum\limits_{x'=1}^xq_{x'}^+ +
\sum\limits_{x=1}^N \sum\limits_{x'=1}^x
\sum\limits_{x''=1}^xq_{x'}^+q_{x''}^+ \\[0,2cm]
{} &=& NE_0^2+ 2E_0\sum\limits_{x=1}^N (N-x+1) q_{x}^+ +
\sum\limits_{x'=1}^N \sum\limits_{x''=1}^N
[N+1-\max(x',x'')]q_{x'}^+q_{x''}^+ \,\, .
\end{array}
\end{equation}
Using the identity
\begin{equation}
\max(x',x'') = {|x'-x''| \over 2} + {{x' + x''} \over 2}
\end{equation}
and rearranging terms leads to
\begin{equation}\label{E3}
\begin{array}{lcl}
\sum\limits_{x=1}^N E_x^2 &=& \,\, N\left[E_0^2+ 2E_0
\sum\limits_{x=1}^N q_{x}^+ +
(\sum\limits_{x=1}^N q_{x}^+)^2\right] - \left(2E_0 + \sum\limits_{x=1}^N
q_{x}^+\right)\sum\limits_{x=1}^N (x-1)q_{x}^+ \\[3mm]
{}&{}& \,\,\,\, -\,\, \sum\limits_{x'=1}^N \sum\limits_{x''=1}^N
{|x'-x''| \over 2} q_{x'}^+q_{x''}^+ \,\, .
\end{array}
\end{equation}
The value of the constant background field $E_0$ is obtained by
setting
\begin{equation}
2\pi^2 \sum_{x=1}^N E_x^2 = \pi^2 \sum_{x,x'=1}^N q_x^+ q_{x'}^+ U_{\mbox{\tiny
P,R}}^+(x-x')\,\, .
\end{equation}
In the large $N$ limit, this amounts to comparing the expression in (\ref{E3})
with the Coulomb interaction part of (\ref{HPBCf}) and (\ref{HRBC}) [after
insertion of (\ref{UR})]. This gives
\begin{equation}\label{E0}
E_0 = \left\{ \begin{array}{r@{\quad \quad}l} m + \frac{\pi_N^{12} +
\pi_N^{21}}{2\pi}\,\, , & \mbox{for} \, \, {\cal H}_{\mbox{\tiny
P}}\,\,\mbox{and}\,\,
{\cal H}_{\mbox{\tiny AP}}\,\, , \\
-{1 \over 2} \sum\limits_{x=1}^N q_x^+\,\, , & \mbox{for} \, \, {\cal
H}_{\mbox{\tiny R}}\,\, ,
\end{array} \right.
\end{equation}
and the Hamiltonians read hence
\begin{equation}\label{HE}
\begin{array}{rcl}
{\cal H}_{\mbox{\tiny P,AP}} &=& \hspace{0,5cm} 2\pi^2\sum\limits_{x=1}^N E_x^2
+
{\sqrt{2} \over 8} \pi^2 \sum\limits_{x,x'=1}^N (3-2\sqrt{2})^{d(x,x')}
q_x^-q_{x'}^- \\[2mm]
{}&{}& +\,\, {{8\pi^2} \over N} (n+{1 \over 4} \sum\limits_{x=1}^N q_x^-
+ \sum\limits_{x=1}^N {\pi_{(x,1)}\over {2\pi}} +
\frac{1}{2}\delta_{\mbox{\tiny AP}})^2
\\[3mm]
{\cal H}_{\mbox{\tiny R}}  &=& \hspace{0,5cm} 2\pi^2\sum\limits_{x=1}^N E_x^2 +
{\sqrt{2} \over 8} \pi^2\, \sum\limits_{x,x'=1}^N\, (3-2\sqrt{2})^{d(x,x')}
\,\sigma (x,x')\, q_x^-q_{x'}^-
\end{array}
\end{equation}
where
\begin{equation}
\delta_{\mbox{\tiny AP}} = \left\{ \begin{array}{r@{\quad \quad}l} 0\, , &
\hspace{0,7cm}\mbox{for} \,\,\mbox{PBC} \,\, , \\
1 \, , & \hspace{0,7cm} \mbox{for} \,\,\mbox{APBC} \,\, ,
\end{array} \right.
\end{equation}
with, from (\ref{ZP}) and (\ref{ZR}),
\begin{equation}\label{constr}
\begin{array}{lcll}
\sum\limits_{x=1}^N q_x^+ &=& \hspace{0,5cm} 0 &
\hspace{0,7cm}\mbox{for} \,\,{\cal H}_{\mbox{\tiny P}}\,\,\mbox{and}\,\, {\cal
H}_{\mbox{\tiny AP}}\,\, , \\
\sum\limits_{x=1}^N q_x^+ \bmod 2 &=& \frac{\pi_N^{12} + \pi_N^{21}}{\pi}
\bmod 2 & \hspace{0,7cm}\mbox{for} \,\,{\cal H}_{\mbox{\tiny R}} \,\, .
\end{array}
\end{equation}

Having established the Hamiltonians for the different
boundary conditions, we are now ready to determine those properties of their
ground
state configurations that are sufficient to calculate the typical energy
difference between the ground state energies for $N\to \infty$.

\setcounter{equation}{0}
\section {The ground states for the different boundary conditions}

We summarise the problem to which the preceding sections have led. Each of the
three expressions (\ref{HE}) should now be minimised with respect to the
variables $n$, $\{E_x\}$, and $\{q_x^-\}$. The $\{E_x\}$ are defined in terms
of $m$ and $\{q_x\}$ by (\ref{E1}) and (\ref{E0}), and the $\{q_x^+\}$ and
$\{q_x^-\}$ are defined in terms of the original charges $\{q_r\}$ by
(\ref{q}). The variables $m$ and $n$ are integers, and the $\{q_r\}$ are
half-integer or integer according to whether the plaquette $r$ is or is not
frustrated.\\[2mm]

The ground states of the Hamiltonians (\ref{HE})
possess in the large $N$ limit the following properties:
{\em \begin{enumerate}
\item The electric field satisfies $|E_x|\laq \frac{1}{2}$ for $x=$ 1,\ldots,N.
The constant background field takes values $|E_0|\laq \frac{1}{2}$.
\item The charges $q_{(x,1)}$, $q_{(x,2)}$ take the values
$0$,$\pm\frac{1}{2}$.
\item On doubly frustrated columns, the charges $q_{(x,1)}$ and $q_{(x,2)}$ are
equal if
and only if $|E_{x-1}|=\frac{1}{2}$.
\end{enumerate}
}

These properties of the ground states, for $N\to \infty$, are proved in
Appendix A. Property {\em 1}\/ indicates that the system's ground state is
within the set of states that minimise the electric field (i.e., the Coulomb)
energy. For PBC and APBC charge neutrality implies $E_N=E_0$ (cf.~equation
(\ref{E1})), whereas for RBC antiperiodicity of the strong interaction
potential leads to $E_N=-E_0=\frac{1}{2}\sum\limits_{x=1}^N q_x^+$ (from
equations (\ref{E1}) and (\ref{E0})). Let us just note here that it was
conjectured in \cite{Vspg} that, in the ground state
of frustrated $XY$ spin systems, the chiralities (charges) $q_r$ are likely
to be zero on nonfrustrated plaquettes and take the values
$\pm\frac{1}{2}$ on frustrated plaquettes. The above property {\em 2}\/ shows
that, for large $N$, this is
{\em indeed}\/ the case for the tube lattice.\\[2mm]

We will now construct the set of states that have properties {\em 1}\/ and {\em
2}\/: We start placing charges $q_{(x,y)}=0,\pm\frac{1}{2}$ successively from
$x=1$ to $x=N$, minimising the local electric field energy density $E_x^2$ at
each step and knowing that it changes by a
half-integer amount (with $q_x^+=\pm\frac{1}{2}$) on a column with one
frustrated plaquette and by an integer amount (with $q_x^+=0$ for $E_{x-1}=0$
and $q_x^+=0,\pm 1$ otherwise) on a column with both plaquettes frustrated (see
figure 2).

Having thus obtained a state that has properties {\em 1}\/ and {\em 2}\/, we
see that it is always possible to partition the nonzero charges $q_{(x,y)}$
into dipoles as in figure 2, grouping together two successive charges of
opposite sign along the $x$-axis such that outside the dipoles the electric
field is zero. [There is one exceptional case (see figure 3): for ${\cal
H}_{\mbox{\tiny R}}$ and $E_0 = {1 \over 2}$ the last and the first charge
placed are not part of any dipole, but of the same sign to take
account of the antiperiodicity of the potential $U^+_{\mbox{\tiny R}}(X) = -
U^+_{\mbox{\tiny R}}(X+N)$, respectively $E_N=-E_0$, as mentioned above.] As
one sees from figure 2, e.g.~from the dipole containing charges on the columns
$x_1$ and $x_1'$,
dipole reversals do not change the Coulomb (i.e., electric field) energy of the
system.
So, the ground state of the system is found within a set consisting of chains
of dipoles, satisfying properties {\em 1}\/~and {\em 2}\/, degenerate in
Coulomb energy. The possibility of columns with two identical charges leads to
the partition into dipoles not being unique. Property {\em 3}\/, however,
introduces a further constraint on the set among which one finds the ground
state. Furthermore, one can easily convince oneself that this latter property
implies that one can reach {\em every}\/ state that satisfies properties {\em
1}\/~to {\em 3}\/ from {\em any}\/ other such state, by reversals of dipoles,
for {\em any}\/ given partition. In particular, the ground state of the system
differs, for large $N$ from a state as constructed above (with the additional
constraint from property {\em 3}\/) by a reversal of dipoles.
\begin{figure}
\setlength{\unitlength}{1cm}
\begin{picture}(14,7)
\thicklines
\put(1.5,4){\line(1,0){12}}
\put(1.5,5){\line(1,0){12}}
\put(1.5,6){\line(1,0){0.06175}}
\put(13.43825,6){\line(1,0){0.06175}}
\multiput(1.81175,6)(0.25,0){47}{\line(1,0){0.125}}
\multiput(2,4)(1,0){12}{\line(0,1){2}}
\put(4.3,5.4){\vector(-2,-1){1.6}}
\put(6.5,4.65){\vector(0,1){0.65}}
\put(7.65,4.65){\vector(1,1){0.65}}
\put(10.3,5.4){\vector(-2,-1){1.6}}
\put(12.35,4.65){\vector(-1,1){0.65}}
\put(13.5435,5.152){\vector(-3,1){0.8}}
\thinlines
\put(0.75,4.25){\vector(0,1){0.5}}
\put(0.75,4){\makebox(0,0){$y$}}
\put(11.25,3.5){\vector(1,0){1.5}}
\put(11,3.5){\makebox(0,0){$x$}}
\put(2.5,4.5){\makebox(0,0){+}}
\put(4.5,5.5){\makebox(0,0){-}}
\put(6.5,4.5){\makebox(0,0){-}}
\put(6.5,5.5){\makebox(0,0){+}}
\put(7.5,4.5){\makebox(0,0){-}}
\put(8.5,4.5){\makebox(0,0){+}}
\put(8.5,5.5){\makebox(0,0){+}}
\put(10.5,5.5){\makebox(0,0){-}}
\put(11.5,5.5){\makebox(0,0){+}}
\put(12.5,4.5){\makebox(0,0){-}}
\put(12.5,5.5){\makebox(0,0){+}}
\thicklines
\put(1.45,1.5){\line(1,0){1.05}}
\put(2.5,2){\line(1,0){2}}
\put(4.5,1.5){\line(1,0){3}}
\put(7.5,1){\line(1,0){1}}
\put(8.5,2){\line(1,0){2}}
\put(10.5,1.5){\line(1,0){1}}
\put(11.5,2){\line(1,0){2}}
\put(2.5,1.5){\line(0,1){0.5}}
\put(4.5,1.5){\line(0,1){0.5}}
\put(7.5,1){\line(0,1){0.5}}
\put(8.5,1){\line(0,1){1}}
\put(10.5,1.5){\line(0,1){0.5}}
\put(11.5,1.5){\line(0,1){0.5}}
\put(2.5,1.25){\makebox(0,0){$x_1$}}
\put(4.5,1.25){\makebox(0,0){$x_1'$}}
\put(1,2){\makebox(0,0){$+\frac{1}{2}$}}
\put(1,1){\makebox(0,0){$-\frac{1}{2}$}}
\thinlines
\put(1.45,1){\line(1,0){0.1}}
\put(1.45,2){\line(1,0){0.1}}
\put(1.5,0.5){\vector(0,1){2}}
\put(1.45,1.5){\vector(1,0){12.1}}
\put(1.3,2.8){\makebox(0,0){$E_x$}}
\put(13.75,1.25){\makebox(0,0){$x$}}
\multiput(2.5,1.45)(1,0){11}{\line(0,1){0.1}}
\end{picture}

{\footnotesize {\noindent\bf\footnotesize Figure 2:} {\sl\footnotesize
Construction of a state that possesses properties 1 and 2.} We
start placing charges $q_{(x,y)}$ from the left to the right. We proceed in a
way that the local electric field stays as close to $0$ as possible. The signs
indicate
plaquettes with charges $\pm\frac{1}{2}$. Plaquettes with no sign are
charge-free (unfrustrated).}
\end{figure}
\begin{figure}
\setlength{\unitlength}{1cm}
\begin{picture}(14,7)
\thicklines
\put(1.5,4){\line(1,0){12}}
\put(1.5,5){\line(1,0){12}}
\put(1.5,6){\line(1,0){0.06175}}
\put(13.43825,6){\line(1,0){0.06175}}
\multiput(1.81175,6)(0.25,0){47}{\line(1,0){0.125}}
\multiput(2,4)(1,0){12}{\line(0,1){2}}
\thinlines
\put(0.75,4.25){\vector(0,1){0.5}}
\put(0.75,4){\makebox(0,0){$y$}}
\put(11.25,3.5){\vector(1,0){1.5}}
\put(11,3.5){\makebox(0,0){$x$}}
\put(2.5,4.5){\makebox(0,0){+}}
\put(3.5,5.5){\makebox(0,0){-}}
\put(5.5,4.5){\makebox(0,0){+}}
\put(5.5,5.5){\makebox(0,0){-}}
\put(6.5,4.5){\makebox(0,0){-}}
\put(9.5,5.5){\makebox(0,0){-}}
\put(10.5,4.5){\makebox(0,0){+}}
\put(11.5,5.5){\makebox(0,0){-}}
\put(12.5,4.5){\makebox(0,0){-}}
\put(12.5,5.5){\makebox(0,0){+}}
\multiput(8.5,3.775)(0,0.1){24}{\line(0,1){0.05}}
\put(8.5,3.6){\makebox(0,0){$N$}}
\thicklines
\put(1.45,1.5){\line(1,0){1.05}}
\put(2.5,2){\line(1,0){1}}
\put(3.5,1.5){\line(1,0){3}}
\put(6.5,1){\line(1,0){2}}
\put(8.5,2){\line(1,0){1}}
\put(9.5,1.5){\line(1,0){1}}
\put(10.5,2){\line(1,0){1}}
\put(11.5,1.5){\line(1,0){2}}
\put(2.5,1.5){\line(0,1){0.5}}
\put(3.5,1.5){\line(0,1){0.5}}
\put(6.5,1){\line(0,1){0.5}}
\put(9.5,1.5){\line(0,1){0.5}}
\put(10.5,1.5){\line(0,1){0.5}}
\put(11.5,1.5){\line(0,1){0.5}}
\put(8.5,0.75){\makebox(0,0){$N$}}
\put(1,2){\makebox(0,0){$+\frac{1}{2}$}}
\put(1,1){\makebox(0,0){$-\frac{1}{2}$}}
\put(3.35,5.35){\vector(-1,-1){0.65}}
\put(5.5,5.35){\vector(0,-1){0.65}}
\put(11.35,5.35){\vector(-1,-1){0.65}}
\put(12.5,4.65){\vector(0,1){0.65}}
\thinlines
\put(1.45,1){\line(1,0){0.1}}
\put(1.45,2){\line(1,0){0.1}}
\put(1.5,0.5){\vector(0,1){2}}
\put(1.45,1.5){\vector(1,0){12.1}}
\put(1.3,2.8){\makebox(0,0){$E_x$}}
\put(13.75,1.25){\makebox(0,0){$x$}}
\multiput(2.5,1.45)(1,0){11}{\line(0,1){0.1}}
\multiput(8.5,1.025)(0,0.1){9}{\line(0,1){0.05}}
\end{picture}

{\footnotesize {\noindent\bf\footnotesize Figure 3:} {\sl\footnotesize
Example of a state with properties 1~and 2~for RBC, while $E_0=\frac{1}{2}$.}
For RBC, proceeding in the
construction as described in the text, the last charge
to be placed is of the same sign as the first. This is obvious from
equation (\ref{E0}), $E_0 = - \frac{1}{2}\sum\limits_{x=1}^N\, q_x^+$, which
indicates that, in general, there is a surplus of two charges $q_{(x,y)}$ with
opposite sign to $E_0=\pm\frac{1}{2}$ (here $E_0=\frac{1}{2}$), to take account
of the fact that for RBC the potentials $U^{\pm}_{\mbox{\tiny R}}$ are
antiperiodic (see (\ref{URS})). But still all charges, but the first and the
last one, can be grouped into
dipoles as announced in the text.}
\end{figure}

The exact ground state configuration remains unknown, but we know that it
minimises the Coulomb energy independently of the other energies involved and
have characterised the set of Coulomb energy ground states.
[Let us just point out here that, when passing from PBC to RBC, one conserves
by virtue of (\ref{E0}) and (\ref{constr}) the property $E_0=0$ or $E_0\ne 0$
for the ground states at both boundary conditions, due to the fact that
$\frac{\pi^{12}_N+\pi^{21}_N}{2\pi}$ stays the same [see also Appendix A,
especially equation (\ref{E0g})]. So the Coulomb energy stays indeed the same.]
The only remaining degrees of freedom in this set are the directions of single
dipoles: The degeneracy is lifted by the other terms in the Hamiltonians of
equation (\ref{HE}), which fix these directions. The amount of Coulomb energy
of the system with the different boundary conditions being the same, it is the
effect of these {\em other}\/ terms that give rise to the difference between
ground state energies when one varies the boundary conditions. We address this
issue in detail in the following section. In spite of the fact that we ignore
the exact ground state configurations, the above properties s!
uffice to analyse and determine th
e ground state energy differences for $N\to \infty$.

\setcounter{equation}{0}
\section {Boundary conditions and ground state energy differences}

\subsection {Generalities}

There is a general relation between the
finite size scaling of the energy difference, $\Delta E^{(N)}
\stackrel{N\to \infty}{\simeq} J N^{-y}$ (where $J$ is the energy scale), of
the ground states of a system for different boundary conditions and the
corresponding correlation length, $\xi(T)$, at a finite temperature $T$: The
correlation length $\xi(T)$
is set by $\Delta E^{(\xi)} \sim k_{\mbox{\tiny B}}T$, hence
\begin{equation}\label{xi}
\xi (T) \sim \left( {{k_{\mbox{\tiny B}}T} \over J} \right) ^{-{1 \over y}}\,\,
, \qquad T\to 0\,\, .
\end{equation}
[Let us just note here that the energy difference may be either concentrated in
a domain wall or associated with a continuous variation of the order
parameter.]

In the Villain model, we may study the spin-spin correlation and the
chirality-chirality correlation by applying APBC and RBC \cite{KaT,NHM,NH}. So
we have to calculate
\begin{equation}\label{Ediff}
\begin{array}{lcc}
\Delta E_{\mbox{\tiny AP}}^{(N)} &=& E_{\mbox{\tiny AP}}^{(N)} - E_{\mbox{\tiny
P}}^{(N)} \,\, ,\qquad N\to \infty\,\, ,\\
\Delta E_{\mbox{\tiny R}}^{(N)} &=& E_{\mbox{\tiny R}}^{(N)} - E_{\mbox{\tiny
P}}^{(N)}\,\, ,\qquad N\to \infty\,\, ,
\end{array}
\end{equation}
where $E_{\mbox{\tiny P}}^{(N)}$, $E_{\mbox{\tiny AP}}^{(N)}$, and
$E_{\mbox{\tiny R}}^{(N)}$ are the ground state energies under P, AP, and R
boundary conditions, respectively.

In this section and in Appendix B, we will call a
dipole with charges $q_{(x,y)}$ and $q_{(x',y')}$ {\em slanted}\/, if $y \ne
y'$, and {\em horizontal}\/, if $y=y'$. [With this definition, the slanted ones
include the vertical dipoles.] As the
probability for a plaquette to be frustrated is the same for all plaquettes,
a dipole is as likely to be slanted as horizontal.

Furthermore, writing $U^-$ for $U^-_{\mbox{\tiny P}}$ at PBC (\ref{U-P}) and
for $U^-_{\mbox{\tiny R}}$ at
RBC (\ref{UR}), the weak (dipolar) interaction $U^-(\ell)$ has the property
\begin{equation}
U^-(\ell ) > \sum_{{\ell ' = \ell +1}}^\infty U^- (\ell ')
\end{equation}
for all boundary conditions, so that we may approximate its effect by
restricting the interactions of
each nonzero charge $q_x^-$ to those with its two nonvanishing neighbouring
charges. Upon renumbering the nonzero charges $q_x^-$ on the frustrated columns
by
a new index $s = 1,2,\ldots,N_c$ (where $N_c$ is the total number of the
frustrated
columns with nonzero $q_x^-$), we can finally rewrite the effective weak
(dipolar) interaction as
\begin{equation}\label{Ueff}
\sum\limits_{x,x'}\, q_x^-q_{x'}^-\, U^-(x-x') = \sum_{{s=1}}^{N_c} U_s
q_s^-q_{s+1}^- + U(0) \sum_{{s=1}}^{N_c} (q_s^-)^2 \,\, ,
\end{equation}
where $q_{N_c+1} \equiv q_1$ for PBC/APBC and $q_{N_c+1} \equiv -q_1$
for RBC. The charges $q_s^-$ take the values $\pm
{1\over 2}$ and $\pm 1$,
and the $U_s$ are independent quenched random interaction constants.
Since in the set of states that we consider the Coulomb energy is boundary
condition independent, we will deduce the energy differences, $\Delta
E_{\mbox{\tiny AP}}^{(N)}$ and $\Delta E_{\mbox{\tiny R}}^{(N)}$ ($N \to
\infty$), from (\ref{Ueff}) and from the global spin wave term in (\ref{HE}).
The large $N$ limit is selfunderstood in what follows.

\subsection {Antiperiodic boundary conditions}\label{APB}

The actual ground state minimises the second and third term in ${\cal
H}_{\mbox{\tiny P}}$ and ${\cal H}_{\mbox{\tiny AP}}$, equation (\ref{HE}),
within
the space of degenerate ground states of the Coulomb energy, characterised in
the preceding section. The second term, rewritten in equation
(\ref{Ueff}), is the weak interaction between the $q_x^-$. In Appendix B
we show that its lowest-lying excitation lies typically an
energy amount $\sim JN^{-y_c}$ above the ground state, and
determine the exponent $y_c=\log_{\frac{8}{3}}(3+2\sqrt{2}) = 1.7972\ldots$ .
The third term is
\begin{equation}\label{cond0}
\begin{array}{lcl}
\frac{8\pi^2}{N}\left(n+\frac{1}{2}\sum\limits_x q_{(x,1)} +
\sum\limits_x \frac{\pi_{(x,1)}}{2\pi}\right)^2 &{}&\mbox{in}\,\, {\cal
H}_{\mbox{\tiny P}}\,\, ,
\\[3mm]
\frac{8\pi^2}{N}\left(n+\frac{1}{2}\sum\limits_x q_{(x,1)} +
\sum\limits_x \frac{\pi_{(x,1)}}{2\pi} + \frac{1}{2}\right)^2
&{}&\mbox{in}\,\, {\cal H}_{\mbox{\tiny AP}}\,\, ,
\end{array}
\end{equation}
where we have used the neutrality condition to write $\sum\limits_x
q_x^- = 2\, \sum\limits_x q_{(x,1)}$. The terms in (\ref{cond0}), of order
$N^{-1}$,
have their origin in a global spin wave, i.e., of wave length $> N$, which
helps the system to adjust to its boundary conditions when there is a
rotational mismatch (cf.~\cite{NHM,NH}).

One might wonder if it is always possible, by choosing $n$ properly, that the
terms in (\ref{cond0}) vanish in the ground state in
${\cal H}_{\mbox{\tiny P}}$ and/or ${\cal H}_{\mbox{\tiny AP}}$.  As $n+
\sum\limits_{x=1}^N
{\pi_{(x,1)} \over
{2\pi}} \in \{\,0, \pm {1 \over 2}, \pm 1,\ldots\,\}$, this depends
obviously on the number of nonzero charges $q_{(x,1)}$. Hence we have
two cases:
\begin{itemize}
\item[{\em (i)}] the number of frustrated plaquettes $(x,1)$ is
even;
\item[{\em (ii)}] the number of frustrated plaquettes $(x,1)$ is odd.
\end{itemize}
Correspondingly:
\begin{equation}\label{cond}
\begin{array}{lcl}
{1 \over 2} \sum\limits_{x=1}^N q_{(x,1)} \in \{\,0, \pm
{1 \over 2}, \pm 1, \ldots\,\}&{}& \quad\mbox{case {\em (i)}\/}\,\, ,\\[2mm]
{1 \over 2} \sum\limits_{x=1}^N q_{(x,1)} \in \{\, \pm
{1 \over 4}, \pm\frac{3}{4}, \ldots\,\}&{}& \quad\mbox{case {\em (ii)}}\,\, .
\end{array}
\end{equation}
We investigate these cases further.
\vspace{0.3cm}

\noindent {\em (i) Even number of frustrated plaquettes $(x,1)$}\/:
Given the set of $\pi_{(x,1)}$
and possibly reversing a sequence of dipoles as in Appendix B to get\,
$\frac{1}{2}\sum\limits_x q_{(x,1)} + \sum\limits_x
\frac{\pi_{(x,1)}}{2\pi}$\, integer for PBC and halfinteger for APBC
or vice versa, the terms in
(\ref{cond0}) vanish for both Hamiltonians by a proper choice of $n$. As
there is a difference of $\frac{1}{2}$ in the term in parentheses in
(\ref{cond0}), the ground states of ${\cal H}_{\mbox{\tiny P}}$ and ${\cal
H}_{\mbox{\tiny AP}}$
differ by a reversal of a sequence of dipoles containing an odd number
of $q_{(x,1)}\ne 0$, i.e., a sequence of dipoles among which an odd number is
slanted. Hence we obtain, reinserting $J$,
\begin{equation}\label{APi}
\Delta E_{\mbox{\tiny AP}}^{(N)} \sim \pm JN^{-y_c} , \quad \qquad \mbox{case
{\em (i)}\/} \,\, ,
\end{equation}
where the sign indicates that either state, at PBC or APBC, has the
lower ground state energy.
\vspace{0.2cm}

\noindent {\em (ii) Odd number of frustrated plaquettes $(x,1)$}\/:
Here, from (\ref{cond}), the
terms (\ref{cond0}) in ${\cal H}_{\mbox{\tiny P}}$ and ${\cal H}_{\mbox{\tiny
AP}}$ are always
nonzero and the optimal value of $n$ will give an energy
$J\frac{\pi^2}{4N}$ irrespective of the directions of the dipoles. These will
thus be determined by the weak interaction only and be the same
for both boundary conditions. The energy difference is hence
\begin{equation}\label{APii}
\Delta E_{\mbox{\tiny AP}}^{(N)} = 0, \quad \qquad \mbox{case {\em (ii)}} \,\/
,
\end{equation}
in this case.

\subsection {Reflecting boundary conditions}

The ground states at PBC and RBC minimise the second and the third term in
${\cal H}_{\mbox{\tiny P}}$ and the second term in ${\cal H}_{\mbox{\tiny R}}$,
equation (\ref{HE}), within the set of states characterised in the preceding
section. So, again, one has to
distinguish between an even and odd number of $q_{(x,1)}\ne 0$, when
calculating the typical ground state energy difference.
\vspace{3mm}

\noindent {\em (i) Even number of frustrated plaquettes $(x,1)$}\/: We saw
in section \ref{APB} that the term for PBC in (\ref{cond0}) vanishes in the
ground state of ${\cal H}_{\mbox{\tiny P}}$ and that, for half of the samples,
one finds the ground state at PBC by minimising the weak (dipolar) interaction.
If one changes to RBC for {\em this}\/ half of the samples, while keeping the
PBC ground state configuration, one will almost always be able to lower the
energy [Note that the sign of the bond that passes column $N$, which is almost
never the weakest one, has changed (!)]: one just has to reverse a sequence of
dipoles starting at column $N$ in such a way that only one
of the weakest bonds is broken. For the {\em other}\/ half of the samples, the
cancellation of the global spin wave term implies a reversal of a sequence of
dipoles in the
configuration {\em after}\/ minimisation of the dipolar interaction. For these
latter samples, when one changes to RBC, one has to reverse again the same
sequence that was reversed to get the ground state at PBC. In {\em both}\/
cases, this leads to
\begin{equation}\label{Ri}
\Delta E_{\mbox{\tiny R}}^{(N)} \sim \pm J N^{-y_c} , \quad \qquad \mbox{case
{\em (i)}\/} \,\, ,
\end{equation}
for the difference in ground state energies. The minus sign applies
for the first half of the samples, the plus for the second half.
\vspace{2mm}

\noindent (ii) {\em Odd number of frustrated plaquettes $(x,1)$}\/: The global
spin wave term never vanishes in PBC, so that
\begin{equation}\label{Rii}
\Delta E_{\mbox{\tiny R}}^{(N)} \simeq \frac{\pi^2}{4} J N^{-1} \, , \quad
\qquad \mbox{case {\em (ii)}}\, \, ,
\end{equation}
neglecting a possible contribution of order $J N^{-y_c}$.

\setcounter{equation}{0}
\section{Conclusion}

We have studied the $XY$ spin glass with $\pm J$ bonds on a tube
lattice. This system has both a continuous (spin) and a
discrete (chiral) symmetry, and hence two order parameters play a
r\^ole. Our purpose was to determine the divergence, for $T\to 0$, of
the chiral and the spin correlation lengths, via the finite size scaling
of the ground state energy differences under different boundary
conditions. In the presence of two symmetries, the usual
single-symmetry relation between the finite size scaling exponents of the
ground state energy difference
and the correlation length has to be extended in a nontrivial way.
Nevertheless, the spin correlation length exponent $y_s$ (see equation
(\ref{xi}))
is given by the energy difference when one passes from periodic to
antiperiodic boundary conditions, namely
$\overline{(\Delta E_{\mbox{\tiny AP}}^{(N)})^2}^\frac{1}{2} \sim J N^{-y_s}$.
New boundary
condi\-tions, reflecting ones, were introduced \cite{KaT}
to determine the chirality correlation length expo\-nent $y_c$.

The difficulty in performing such an analysis on a general $N\times M$
lattice is that one does
not know how to construct the ground states of the disordered systems. The
tube lattice, of this work, however, just as the ladder lattice studied earlier
\cite{NHM}, allows for a precise
theoretical analysis of this relation. In contrast to the ladder lattice, the
tube lattice still has long-range interactions between its chiralities, and is
therefore closer to a two-dimensional system.

We first apply the well-known transformation
\cite{Vspg,NH} of the $XY$ spin glass into a Coulomb gas, a system of chiral
variables (also called charges). The
resulting effective Hamiltonian can be cast in the form (\ref{HE}) where
it is the sum of three terms:
{\em \begin{itemize}
\item[(i)] A one-dimensional Coulomb interaction, linearly increasing
with distance, bet\-ween charges $q_1^+,$ $q_2^+, \ldots, q_N^+$; in
{\rm (\ref{HE})}, this term has been expressed as the volume sum of the energy
density of the electric field $E_x$.
\item[(ii)] A ``dipolar" interaction that decreases exponentially with distance
between the $q_1^-,q_2^-,\ldots,q_N^-$.
\item[(iii)] The energy of a spin wave needed to match PBC or APBC (but
absent under RBC), and whose wavenumber depends on the total electric
dipole moment.
\end{itemize}}
The third term disappears in the thermodynamic limit. Its relevance for a
finite size scaling analysis was first pointed out by Fisher, Tokuyasu
and Young \cite{FTY}.
Moreover, the three terms are, on the one hand, coupled by {\em local}\/
constraints, that link the allowed values of $q_x^+$ and  $q_x^-$ with the
fixed values of the
ferromagnetic or antiferromagnetic bonds $\pi_{ij}$ between the spins on
the lattice, and, on the other hand, by a {\em global}\/ constraint on the
total charge (zero for PBC and APBC, and even or uneven for RBC). Taking these
constraints into
account, we identify the low-lying excitations of the three terms,
respectively:
{\em \begin{itemize}
\item[(i)] Coulomb excitations that cost an energy of order $J$.
\item[(ii)] Chiral excitations, obtained by reversing a sequence of chiral
variables, that cost an energy $\sim J N^{-y_c}$ with
$y_c=1.7972\ldots$ .
\item[(iii)] Global spin waves that cost an energy $\sim J N^{-1}$.
\end{itemize}}
The $\pm J$ $XY$ spin glass on the ladder lattice \cite{NHM} consists of both
interactions {\em (ii)}\/ and {\em (iii)}\/. Due to the additional long-range
interaction {\em (i)}\/, the tube is closer to the two-dimensional model.

In spite of the number of interactions
in competition, we were able to characterise and delimit the set of charge
configurations, within which lies the ground state. In the configurations
contained in this set, the charges take the values $\pm {1 \over 2}$ on the
frustrated plaquettes and zero on the others, and form a chain of dipoles.

We now give a summary of our results, and recall numerical results for
comparison. When changing boundary conditions from PBC to APBC, or RBC, it is
the excitations {\em (ii)}\/ and {\em (iii)}\/ that give {\em both}\/ energy
differences, $\Delta E_{\mbox{\tiny AP}}^{(N)}$ and $\Delta E_{\mbox{\tiny
R}}^{(N)}$. This implies the same conclusions as in \cite{NHM}: Firstly, the
ground state obtained with P boundary conditions can adjust to AP boundary
conditions {\em via a chiral excitation}\/, so that
\begin{equation}\label{EAPres}
\overline{(\Delta E_{\mbox{\tiny AP}}^{(N)})^2}^\frac{1}{2} \sim J N^{-y_c}\,\,
, \qquad y_c=1.7972\ldots\,\, .
\end{equation}
The last equation contains {\em no reference to spin waves}\/ and means that
$y_s=y_c$.
Secondly,  passing from P to R boundary conditions releases a
global spin wave (as was first observed
by Kawamura and Tanemura \cite{KaT} in $d=2$) in half of the
samples, but does not do so in the other half.

In $d=2$, Kawamura and Tanemura performed a numerical analysis of the different
ground state energies of the cosine $XY$ model. They find, as $N\rightarrow
\infty$,
\begin{equation}\label{KaTc}
\begin{array}{lcllcl}
\overline{(\Delta E_{\mbox{\tiny AP}}^{(N)})^2}^\frac{1}{2} &\simeq& a\,
N^{-y_s}\,\, ,&y_s &\approx& 0.84\, , \\
\overline{(\epsilon^{(N)}_{\mbox{\tiny R}} -
\overline{\epsilon^{(N)}_{\mbox{\tiny R}}})^2}^\frac{1}{2} &\simeq&
b\,N^{-y_c} + {\cal O}(N^{-y_s})\,\, , &y_c &\stackrel{<}{\approx}& 0.38\, ,
\end{array}
\end{equation}
where $a$ and $b$ are constants, and a new quantity, namely
\begin{equation}
\epsilon^{(N)}_{\mbox{\tiny R}} \equiv \Delta E_{\mbox{\tiny R}}^{(N)} -
\min(0,\Delta E_{\mbox{\tiny AP}}^{(N)})\,\, ,
\end{equation}
has been introduced. Thus, they get two distinct exponents $y_s$ and $y_c$,
with $y_s > y_c$, and conclude that the chiralities order on a longer scale
than the spin variables.

For the tube, upon collecting our results [equations
(\ref{APi}),(\ref{APii}),(\ref{Ri}) and (\ref{Rii})], we get
\begin{equation}\label{ERres}
\overline{(\epsilon^{(N)}_{\mbox{\tiny R}} -
\overline{\epsilon^{(N)}_{\mbox{\tiny R}}})^2}^\frac{1}{2} =
\frac{\pi^2}{8} J N^{-1} + {\cal O}(N^{-y_c})\,\, ,
\end{equation}
i.e., the R boundary
conditions probe a global spin wave term proportional to $N^{-1}$.
If we now conjecture on the extrapolation of our results to $d=2$,
then we expect for the quantities of equation (\ref{KaTc}) that
$\Delta E_{\mbox{\tiny AP}}$ would yield a chiral exponent $y_c$
as in (\ref{EAPres}) but with a smaller value (since $y_c$ should vanish
at some, still higher, lower critical dimension); and that
$\epsilon_{\mbox{\tiny R}}$ would yield the spin wave exponent
$d-2=0$. Instead, in contrast, Kawamura and Tanemura interpret
their simulation according to (\ref{KaTc}). We expect that simulations
on larger $2d$ systems will confirm our scenario.

\vspace{4.5cm}

\setcounter{equation}{0}
\renewcommand{\theequation}{\mbox{A.\arabic{equation}}}

\begin{appendix}
\noindent {\Large\bf Appendix A}
\vskip 0,5cm
In this appendix, we prove that in the large $N$ limit the ground states of the
system for the
different boundary conditions possess the properties {\em 1}\/ to {\em 3}\/
announced in section III.
In the calculations, we write the expression of the weak (dipolar) interaction
in its
form at PBC/APBC (equation (\ref{U-P})), with again $J=2$. The arguments are
nevertheless readily rewritten for RBC,
including the appropriate factors of $\sigma(x,x')$ (equation (\ref{UR})).
Furthermore, we neglect the global spin wave term ${\cal O}(\frac{1}{N})$
that appears for PBC/APBC. Upon proper choice of $n$, this term contributes
at most $\frac{\pi^2}{2N}$ to the ground state energies at PBC/APBC, which is
small for $N\rightarrow \infty$, in comparison to the other
energies involved.\\[5mm]

In preparation of the proofs of the ground state properties {\em 1}\/ to {\em
3}\/, we show, in a first step, that
\vspace{-0.3cm}
{\em
\begin{itemize}
\item[(i)] in the ground state, the charges $q_x^-$ take values
$|q_x^-|\laq \frac{3}{2}$,
\end{itemize}}\vspace{-0.3cm}
\noindent and, using {\em (i)}\/, in a second step, that
\vspace{-0.3cm}
{\em
\begin{itemize}
\item[(ii)] in the ground state, the charges $q_x^-$ take values
$|q_x^-|\laq 1$.
\end{itemize}\vspace{-0.3cm}
}
${}$

\noindent {\bf Proofs of {\em (i)} and {\em (ii)}:}

\noindent {\em (i) In the ground state, the charges $q_x^-$ take
values such that $|q_x^-|\laq \frac{3}{2}$.\\[0.1cm]}
Let us look at some state with charges $q_x^{-,0}$ such that
\begin{equation}
q \equiv \max\{|q_x^{-,0}|\} \gaq 2\,\, .
\end{equation}
Let $q_{x_0}^{-,0}$ be a charge with $|q_{x_0}^{-,0}|=q$. For reasons of
charge reversal symmetry, we may take $q_{x_0}^{-,0}$ positive without loss
of generality.

Consider now the state in which $q_{x_0}^{-,0}$ is changed into
$q_{x_0}^{-,0}-2$, while all other charges $q_x^{-,0}$ and all
charges $q_x^{+,0}$ are kept unchanged. [Note that, by an appropriate change
of $q_{(x,1)}$ and $q_{(x,2)}$, one can add an arbitrary multiple of 2 to
some charge $q_x^-$ without changing $q_x^+$ (see (\ref{q})).]
The difference in energy $\Delta E$ between the (final) state, with
$q_{x_0}^{-,0}$
changed, and the initial state can readily be calculated. As the
charges $q_x^+$ are unchanged, it comes from the difference in weak
interaction energy only. For $q_{x_0}^{-,0}\gaq 2$, one finds
\begin{equation}
\begin{array}{lcl}
\Delta E &=& \frac{\sqrt{2}}{8}\pi^2\left[(q-2)^2-q^2\right] +
\frac{\sqrt{2}}{8}\pi^2\left[q-(q-2)\right]2\sum\limits_{s>0}(q_{x+s}^{-,0}
+ q_{x-s}^{-,0})(3-2\sqrt{2})^s \\[2mm]
{} &=& \frac{\sqrt{2}}{8}\pi^2 \left[-4(q-1) +
4\sum\limits_{s>0}(q_{x+s}^{-,0}+q_{x-s}^{-,0})(3-2\sqrt{2})^s\right]\,\, .
\end{array}
\end{equation}
We have
\begin{equation}
\left|4\sum\limits_{s>0}(q_{x+s}^{-,0}+q_{x-s}^{-,0})(3-2\sqrt{2})^s\right|
\laq 8q\left|\sum\limits_{s>0}(3-2\sqrt{2})^s\right| \,\, ,
\end{equation}
and thus, summing the geometric series and using $q\gaq 2$,
\begin{equation}
\Delta E \laq \frac{\sqrt{2}}{8}\pi^2 \left[-4(q-1) +
4(\sqrt{2}-1)q\right] < 0\,\, .
\end{equation}
So the final state is lower in energy than the initial state.
Hence a state, in which $q\gaq 2$, is not the ground state.
\vspace{0.1cm}

\noindent {\em (ii) In the ground state, the charges $q_x^-$ take values
such that $|q_x^-|\laq 1$.\\[0.1cm]}
Let us take some state with charges $q_x^{-,0}=0,\pm\frac{1}{2},\pm 1,
\pm\frac{3}{2}$. Be $n$ the number of charges $\pm\frac{3}{2}$, with $n>0$.
Suppose for the moment that $n<N$. There exists hence a sequence of
charges $q_{x_0}^{-,0}$,
$q_{x_0+1}^{-,0}$,$\cdots$, $q_{x_0+n'-1}^{-,0}=\pm\frac{3}{2}$
($1\laq n'\laq n$) that is enclosed between charges $q_x^{-,0}$ of absolute
value $\laq 1$ (i.e., $|q_{x_0-1}^{-,0}|$,$|q_{x_0+n'}^{-,0}|\laq 1$).

Consider now the state in which all charges $q_x^{-,0}=\pm\frac{3}{2}$
in this sequence are replaced by $\mp\frac{1}{2}$, while all other
charges $q_x^{-,0}$ and all charges $q_x^{+,0}$ are kept unchanged.
As in {\em (i)}\/, the difference in energy $\Delta E$ between the (final)
state,
with the changes, and the initial state is due to the difference
in weak (dipolar) interaction energy only. The difference $\Delta E_0$ in weak
interaction energy, coming from the self-interaction terms in the sum, is
\begin{equation}
\Delta E_0 = -2\frac{\sqrt{2}}{8}\pi^2 n'\,\, ,
\end{equation}
the one from the nearest-neighbour interaction terms is of absolute value
\begin{equation}
|\Delta E_1| \laq 2\frac{\sqrt{2}}{8}\pi^2(3-2\sqrt{2})
\left[(n'-1)\frac{9}{4}+2\cdot\frac{3}{2}\cdot 1\right] +
2\frac{\sqrt{2}}{8}\pi^2(3-2\sqrt{2})
\left[(n'-1)\frac{1}{4}+2\cdot\frac{1}{2}\cdot 1\right]\,\, ,
\end{equation}
while the energy difference of all other terms is of absolute value
\begin{equation}
|\Delta E_2| \laq 4\frac{\sqrt{2}}{8}\pi^2\frac{(3-2\sqrt{2})^2}
{1-(3-2\sqrt{2})}\frac{9}{4}n' + 4\frac{\sqrt{2}}{8}\pi^2\frac{(3-2\sqrt{2})^2}
{1-(3-2\sqrt{2})}\frac{3}{4}n'\,\, .
\end{equation}
To obtain the last inequality, we have substituted for all charges $q_x^-$,
but the ones in the sequence, the maximal possible absolute value $\frac{3}{2}$
and taken all terms in the sum to be negative in the initial state and positive
in the final state [which is obviously an upper bound for the
energy change, but impossible to realise]. The overall energy
difference is thus
\begin{equation}\label{dE2}
\begin{array}{lcl}
\Delta E &\laq& -2\frac{\sqrt{2}}{8}\pi^2n' + \frac{\sqrt{2}}
{8}\pi^2(3-2\sqrt{2})\left[5(n'-1)+8\right] + 12\frac{\sqrt{2}}
{8}\pi^2\frac{(3-2\sqrt{2})^2}{1-(3-2\sqrt{2})}n'\\
{}&=& - \frac{\sqrt{2}}{8}\pi^2[n'(29-20\sqrt{2}) - 9 +
6\sqrt{2}] < 0\,\, .
\end{array}
\end{equation}
Again, the energy of the final state is lower than the energy of the initial
state. If $n=N$, a similar reasoning (with $n'=n=N$) leads to the same
conclusion.
This proves {\em (ii)}\/.\\[0.7mm]

\noindent {\bf We are now prepared to show that the ground state has the
properties {\em 1}\/ to {\em 3}\/, announced in section III.}\\[0.5mm]

\noindent {\em 1. In the ground state, the electric field satisfies $|E_x|\laq
\frac{1}{2}$. The constant background field takes values $|E_0|\laq
\frac{1}{2}$.}

\noindent We first note that the electric field is constrained to be of
absolute value $\laq 1$ for the ground state configuration.
This can be seen as follows.
{}From (\ref{HE}), we see that minimisation of the Coulomb interaction part
of the Hamiltonians means minimising the mean square value of the local
electric field $E_x$. This leads for all Hamiltonians to
\begin{equation}\label{E0g}
E_0 = \left\{ \begin{array}{r@{\quad \quad}l} 0\, , & \mbox{if} \, \,
\frac{\pi_N^{12} + \pi_N^{21}}{\pi} \bmod 2 = 0 \,\, , \\
\pm{1 \over 2}\, , & \mbox{if} \, \, \frac{\pi_N^{12} + \pi_N^{21}}{\pi}
\bmod 2 = 1 \,\, ,
\end{array} \right.
\end{equation}
for the constant background field $E_0$. The electric field, $E_x$,
should be locally optimal, that is stay as close to $0$ as possible.
For any given state with given sets of charges $\{q_x^+\}$
and $\{q_x^-\}$, one can go successively through the system, from $x=1$
to $x=N$, changing, whenever necessary, the charges $q_x^+$ in such a way that
the value
of the electric field is bounded in absolute value by $1$ in the final state:
At every column with no
frustrated or two frustrated plaquettes, the change in
electric field can be bounded to be $0,\pm 1$ by an
appropriate change of $q_x^+$, if necessary, and to be
$\pm\frac{1}{2}$ on every column with exactly one frustrated
plaquette. [By the definition of $q_x^+$ and $q_x^-$, equation (\ref{q}), one
can add an
arbitrary multiple of 2 to some charge $q_x^+$ without changing
$q_x^-$, by an appropriate change of $q_{(x,1)}$ and $q_{(x,2)}$.]
Whenever one encounters a column $x_0$, during the above procedure, where the
electric
field jumps to a value $|E_{x_0}|\gaq\frac{3}{2}$, one
changes $q_{x_0}^+$ by the appropriate amount, as well
as the next nonzero charge $q_x^+$, say at $x_0'$, by
the opposite amount (to conserve charge neutrality).
By the definition of the electric field, cf.~equation (\ref{E1}),
every time one changes the charges at $x_0$ and $x_0'$ only, the electric field
remains unchanged on columns $x<x_0$ and $x\gaq x_0'$.
So in the end, while keeping the set of charges $\{q_x^-\}$
fixed, the value of the electric field is bounded in
absolute value by $1$. In particular, the final state is lower in energy
than the initial state.
\vspace{2mm}

Let us now take a charge configuration of the system (with charges
$q_x^{+,0}$, $q_x^{-,0}$) such that the electric field jumps, say at column
$x_0$, to a
value $|E_{x_0}|=1$, and stays at this value until column
$x_0'$ ($>x_0$), where one finds hence the next nonzero
charge $q_x^{+,0}$. Consider the state in which $q_{x_0}^{+,0}$
is changed by an amount of $1$ to some value $|E'_{x_0}|\laq\frac{1}{2}$
and $q_{x_0'}^{+,0}$ by the opposite amount (to conserve charge neutrality),
while keeping the charges $q_x^{+,0}$ (and $q_x^{-,0}$) on all other columns
fixed.
We observe that, from (\ref{E1}), the electric field is unchanged
for $x<x_0$ and $x\gaq x_0'$ and that by the definition of $q_x^+$
and $q_x^-$, the charges $q_{x_0}^{-,0}$ and $q_{x_0'}^{-,0}$
changed. Anyhow, from {\em (ii)}\/, the absolute value of the
charges $q_x^-$ can be taken to be bounded by $1$ in both
states, as otherwise the energy of the state can still be lowered. The
difference
$\Delta E^F$ in strong interaction energy between the final state
and the initial state is
\begin{equation}
\Delta E^F \laq -2\pi^2\frac{3}{4}\,|x_0-x'_0|\,\, ,
\end{equation}
while the difference $\Delta E^f$ in weak interaction energy is
bounded by
\begin{equation}
\begin{array}{lcl}
|\Delta E^f| &\laq& \,\,\,\,\, 2\frac{\sqrt{2}}{8}\pi^2
\left[1+2\left|\sum\limits_{s>0}q_{x_0}^{-,0}(q_{x_0+s}^{-,0}
+q_{x_0-s}^{-,0})(3-2\sqrt{2})^s\right|\right]\\[4mm]
{}&{}&+\,\, 2\frac{\sqrt{2}}{8}\pi^2\left[1+2
\left|\sum\limits_{s>0}q_{x_0'}^{-,0}(q_{x_0'+s}^{-,0}
+q_{x_0'-s}^{-,0})(3-2\sqrt{2})^s\right|\right]\,\, .
\end{array}
\end{equation}
[The factors $2$ in the last inequality stem from the fact that both the energy
of the initial state and the one of the final state enter in the difference.]
Hence we get for the total energy difference
\begin{equation}
\Delta E \laq -2\pi^2\frac{3}{4} + 4\frac{\sqrt{2}}{8}\pi^2
[1+2(\sqrt{2}-1)] < -\frac{\sqrt{2}-1}{2}\pi^2\,\, ,
\end{equation}
i.e., the energy of the final state is lower than the one of the initial state.
Thus $|E_x|\laq\frac{1}{2}$ for the ground state configuration.

\noindent {\em 2. In the ground state, the charges
$q_{(x,1)}$, $q_{(x,2)}$ take the values $0,\pm\frac{1}{2}$.}\/

\noindent From {\em (ii)}\/ and {\em 1}\/, the absolute values of the
charges on the plaquettes in the ground state are bounded by 1. In the case
that $E_{x-1}=0$ in the ground state,
it is easy to see from {\em (ii)}\/ and {\em 1}\/ that the
charges $q_{(x,1)}$, $q_{(x,2)}$ on the plaquettes of column
$x$ are of value $0,\pm\frac{1}{2}$.\\[-0.5cm]

Let us thus consider the
case $|E_{x-1}|=\frac{1}{2}$. If there is at least one frustrated
plaquette on column $x$, it is again obvious from {\em (ii)}\/ and
{\em 1}\/ that $q_{(x,1)}$ and  $q_{(x,2)}$ are of absolute
value $\laq\frac{1}{2}$. Suppose now that there are two nonfrustrated
plaquettes on a certain column $x_0$. From what we have stated at the
beginning of this paragraph, it could be that there is a charge
$\pm 1$ on one of the plaquettes. Let us
furthermore suppose that there is a sequence of columns $x_0$,
$x_0+1$,$\cdots$, $x_0+n-1$ that carry charges $q_x^-$ of
absolute value $1$.\\[-0.5cm]

If the number $n$ of charges $q_x^-$ in the sequence is even, consider
the state in which all the charges $q_{x_0}^-$,$q_{x_0+1}^-$,
$\cdots$,$q_{x_0+n-1}^-$ are changed to $0$ [this is possible as
it conserves charge neutrality, as one convinces oneself with the
help of {\em 1}\/]. The absolute value of the electric field is
the same in both the initial and the final state, so that the difference in
energy is
again due to the weak interaction only:
\begin{equation}
\Delta E \laq -\frac{\sqrt{2}}{8}\pi^2n+ 4\frac{\sqrt{2}}{8}\pi^2n
\frac{3-2\sqrt{2}}{1-(3-2\sqrt{2})} = \frac{\sqrt{2}}{8}\pi^2n
[2\sqrt{2}-3] < 0\,\, ,
\end{equation}
and the energy of the final state is lower than the one of the initial
state.\\[-0.3cm]

In the case that $n$ is odd, one can take $n=1$ and
$|q_{x-1}^-|$,$|q_{x+1}^-|\laq\frac{1}{2}$ without loss of generality, in view
of the preceding paragraph. Suppose first
that one of the charges $q_{x-1}^-$,$q_{x+1}^-$ is of absolute
value $\frac{1}{2}$ in the ground state, say $q_{x+1}^-$ and say
$|q_{(x+1,1)}|=\frac{1}{2}$. Let us compare the energy of this state with
the one in which $q_x^-$ is changed to $0$
and $q_{(x+1,1)}$ replaced by $-q_{(x+1,1)}$ [note that
$q_x^-+q_{(x+1,1)}=-q_{(x+1,1)}$ from {\em 1}\/, so that the
absolute value of the electric field remains unchanged and
charge neutrality is conserved]. Using {\em (ii)}\/, the difference
in energy is
\begin{equation}
\begin{array}{lcl}
\Delta E &\laq& \frac{\sqrt{2}}{8}\pi^2\left[-1+
2(|q_{x-1}^{-,0}q_{x}^{-,0}|
+|q_{x}^{-,0}q_{x+1}^{-,0}|+|q_{x+1}^{-,0}q_{x+2}^{-,0}|)
(3-2\sqrt{2})\right.\\[2mm]
{}&{}& \hspace{1.5cm}+\, 2|-q_{x+1}^{-,0}q_{x+2}^{-,0}|(3-2\sqrt{2})
+2\sum\limits_{s>1}|q_{x}^{-,0}||q_{x-s}^{-,0}+
q_{x+s}^{-,0}|(3-2\sqrt{2})^s\\[2mm]
{}&{}& \hspace{1.5cm}+\, 2\cdot
2\sum\limits_{s>1}|q_{x+1}^{-,0}||q_{x+1-s}^{-,0}
+q_{x+1+s}^{-,0}|
(3-2\sqrt{2})^s\Big]\\[3mm]
{}&\laq& \frac{\sqrt{2}}{8}\pi^2\left[-1+8(3-2\sqrt{2})
\frac{1}{2}+8\frac{(3-2\sqrt{2})^2}{2\sqrt{2}-2}\right]
= \frac{\sqrt{2}}{8}\pi^2\, (-17 + 12\sqrt{2}) < 0\,\, .
\end{array}
\end{equation}
Again the final state is lower in energy than the initial state. Secondly, if
$q_{x-1}^-=q_{x+1}^-=0$,
there are again two possibilities. Either there are only
nonfrustrated plaquettes on the columns $x-1$ and $x+1$, or, on
at least one of them, both plaquettes are frustrated, say at $x+1$.
In the last case, one has $q_x^+=-2q_{(x+1,1)}=-2q_{(x+1,2)}$
from {\em 1}\/; hence changing $q_x^-$ to $0$ and
$q_{(x+1,1)},q_{(x+1,2)}$ to $-q_{(x+1,1)},-q_{(x+1,2)}$
conserves charge neutrality, the absolute value of the
electric field and does not change $q_{x-1}^-=q_{x+1}^-=0$.
Using {\em (ii)}\/, the energy difference between
the final and the initial state is then
\begin{equation}
\begin{array}{lcl}
\Delta E &\laq& \frac{\sqrt{2}}{8}\pi^2\left[-1+
2\sum\limits_{s>1}|q_{x}^{-,0}||q_{x-s}^{-,0}
+q_{x+s}^{-,0}|(3-2\sqrt{2})^s\right]\\[2mm]
{} &\laq& \frac{\sqrt{2}}{8}\pi^2\left[-1+
4\frac{(3-2\sqrt{2})^2}{2\sqrt{2}-2}\right] =
\frac{\sqrt{2}}{8}\pi^2\, (-15 + 10\sqrt{2}) < 0\,\, .
\end{array}
\end{equation}
In the case that $q_{(x-1,1)}=q_{(x-1,2)}=
q_{(x+1,1)}=q_{(x+1,2)}=0$, there is a column $x-s$ or
$x+s$ such that $q_{(x-\sigma,1)}=q_{(x-\sigma,2)}=
q_{(x+\sigma,1)}=q_{(x+\sigma,2)}=0$ for all $1<\sigma<s$
and that one of the charges
$q_{(x-s,1)}$,$q_{(x-s,2)}$,$q_{(x+s,1)}$,$q_{(x+s,2)}$
is nonzero. Considerations analoguous to the ones earlier in this
paragraph lead again to the conclusion that a state in which there is a
nonzero charge on a column with two nonfrustrated plaquettes is not the ground
state.

So in the ground state, the charges $q_{(x,1)}$, $q_{(x,2)}$
take the values $0,\pm\frac{1}{2}$ only.\\[0.3cm]

\noindent {\em 3. In the ground state, the charges $q_{(x,1)}$ and $q_{(x,2)}$
are equal on doubly frustrated columns, if and only if
$|E_{x-1}|=\frac{1}{2}$.}

\noindent Following exactly the same lines as in {\em 2}\/, one shows that a
state in which $q_{(x_0,1)}=-q_{(x_0,2)}$, i.e., $|q_{x_0}^-|=1$, on some
doubly frustrated column $x_0$, while $|E_{x_0-1}|=\frac{1}{2}$, is higher in
energy than the one in which $q_{(x_0,1)}=q_{(x_0,2)}$, i.e., $|q_{x_0}^-=0|$
[the appropriate change of some other charge, as in {\em 2}\/, to conserve
charge neutrality, is tacitly understood]. It is evident from {\em 1}\/ that,
in the ground state, $q_{(x_0,1)}=-q_{(x_0,2)}$ on some doubly frustrated
column $x_0$, if $E_{x_0-1}=0$.
\vspace{2cm}

\setcounter{equation}{0}
\renewcommand{\theequation}{\mbox{B.\arabic{equation}}}
\noindent {\Large\bf Appendix B}
\vskip 0,5cm
We calculate the typical energy change in a tube of length
$N$, in the limit $N \rightarrow \infty$, when a sequence of dipoles
is reversed to adjust to APBC or RBC from PBC, as in section IV. This amount
of energy is related to the typical length of the longest interval that
contains no nonzero charges $q_x^-$ between two such (nonzero) charges. The
reversals that are effectuated in section IV do not necessarily involve
{\it the} longest of these intervals because of the constraints: one must not
brake up a dipole, or in certain cases has to reverse not just {\em any}\/
sequence, but one that
contains an {\em odd}\/ number of slanted dipoles. Nevertheless, these
reversals will
still typically involve intervals that are of the same order as the longest
interval.

Since each plaquette is frustrated with probability $1 \over 2$ independently
of the others, a tube of length $N$ will typically contain $N \over 4$
nonfrustrated columns, $\frac{N}{2}$ with one frustrated plaquette,
and $N \over 4$
where both plaquettes are frustrated. From property {\em 3}\/ of the ground
state configuration, half of the doubly frustrated
columns will typically contain charges that belong to the same dipole and the
other half
charges that belong to two different dipoles. This is due to the fact that at a
given column $E_x = 0$ or $E_x = \pm {1 \over 2}$ (i.e., $q_x^+$ integer or
half-integer) with equal probability; in the first case, one has $q_x^- = \pm
1$, and in the other, $q_x^- = 0$. So, there are more charges $q_x^- = 0$ than
there are nonfrustrated columns.
Since it is typically half of the doubly frustrated columns that give
$q_x^- = 0$, we see that, again typically, ${5 \over 8}N$ of the columns
carry a nonzero charge $q_x^-$, while ${3 \over 8}N$ of the columns are neutral
in $q_x^-$. So the number of intervals between two nonzero charges $q_x^-$,
that contain no other such (nonzero) charge, is ${5 \over 8}N$.
The probability $p(\ell)$ for the two subsequent nonvanishing charges $q_x^-$
to be at a distance $\ell$ (i.e., to be separated
by an interval of $\ell -1$ columns containing no nonzero charge $q_x^-$) is
\begin{equation}
p(\ell ) = \left( {3 \over 8} \right) ^{\ell -1} {5 \over 8}\, , \hspace{2cm}
\ell = 1,2,\ldots\,\, .
\end{equation}

Let $P_N(m)$ be the probability distribution for the length $m$ of the
longest one of these distances. Obviously, $P_N(m)$ equals the probability
that all ${5 \over 8}N$ intervals have length $\ell \laq m$, minus the
probability that they all have length $\ell \laq m - 1$; explicitly
\begin{equation}\label{Pm}
P_N(m) = \left[ \sum_{{\ell = 1}}^m p(\ell ) \right] ^{{5 \over 8}N}
- \left[ \sum_{{\ell = 1}}^{m-1} p(\ell ) \right] ^{{5 \over 8}N}
= \left[ 1 - \left({3 \over 8} \right) ^m \right] ^{{5 \over 8}N}
- \left[ 1 - \left({3 \over 8} \right) ^{m-1} \right] ^{{5 \over 8}N} .
\end{equation}
When $N$ is large, $P_N(m)$ will be peaked around some large value of $m$.
Its scaling form can be obtained if one transforms from $m$ to $m'$ according
to
\begin{equation}\label{m}
m = \gamma \log N + m' \,\, ,
\end{equation}
with $\gamma$ to be determined. Indeed, upon using (\ref{m}) and (\ref{Pm}),
one finds
\begin{equation}
P_N(m) \,\,\stackrel{N\to\infty}{\simeq}\,\, x(1-x^{\frac{5}{3}})\,\, ,\qquad
\mbox{with}\quad x\,=\,e^{-\frac{5}{8}\left(\frac{3}{8}\right)^m\,N}\,\, .
\end{equation}
This shows that $P_N(m)$ is effectively nonzero only for argument values
\begin{equation}
m = \log_{{8 \over 3}}N + {\cal O} (1) \,\, ,
\end{equation}
and that the appropriate scaling limit reads
\begin{equation}
\begin{array}{llcl}
{}&N &\rightarrow& \infty\, , \\
{}&m' &{}& \mbox{finite, fixed}\, , \\
\mbox{and}&\gamma &=& \frac{1}{\log\frac{8}{3}}\, .
\end{array}
\end{equation}

Having thus obtained the typical length
\begin{equation}
m = \frac{\log N}{\log\frac{8}{3}}
\end{equation}
of the longest interval, we are able to determine the typical energy
change of the mentioned dipole reversals. From the effective weak interaction
(\ref{Ueff}) we deduce that the typical energy change due to
such a reversal is of order $c\, U(m)$, where
\begin{equation}
U(m) = (3 - 2 \sqrt{2})^m
\end{equation}
is the energy change when breaking up a bond at distance $m$ and the
constant $c$ is of order unity. The exponent $y_c$ is then obtained from
the defining equation
\begin{equation}
c\, U(0)\, N^{-y_c} = c\, U(m)
\end{equation}
which gives
\begin{equation}
y_c = \log_{8 \over 3}(3+2\sqrt{2}) = 1.7972\ldots
\end{equation}
for the exponent of the chirality-chirality correlation length.

\end{appendix}

\end{document}